# Android Malware Detection Using Parallel Machine Learning Classifiers

Suleiman Y. Yerima, Sakir Sezer
Centre for Secure Information Technologies (CSIT)
Queen's University, Belfast,
Northern Ireland
{s.yerima, s.sezer}@qub.ac.uk

Igor Muttik
Principal Research Architect,
McAfee Labs
London, United Kingdom
igor_muttik@mcafee.com

*Abstract—* Mobile malware has continued to grow at an alarming rate despite on-going mitigation efforts. This has been much more prevalent on Android due to being an open platform that is rapidly overtaking other competing platforms in the mobile smart devices market. Recently, a new generation of Android malware families has emerged with advanced evasion capabilities which make them much more difficult to detect using conventional methods. This paper proposes and investigates a parallel machine learning based classification approach for early detection of Android malware. Using real malware samples and benign applications, a composite classification model is developed from parallel combination of heterogeneous classifiers. The empirical evaluation of the model under different combination schemes demonstrates its efficacy and potential to improve detection accuracy. More importantly, by utilizing several classifiers with diverse characteristics, their strengths can be harnessed not only for enhanced Android malware detection but also quicker white box analysis by means of the more interpretable constituent classifiers.

I. INTRODUCTION

Android malware is growing at an astonishing rate despite the measures currently in use to curtail infection amongst the growing population of Android users worldwide. The primary means of Android app distribution is via app markets, and several unofficial online app stores are emerging alongside the official Google Play app store. In 2012 Google announced the use of 'Bouncer' to screen apps being submitted to its official app store for malicious behavior. In fact, the analysis process of Bouncer, which is based on run-time dynamic analysis, has been demonstrated by Oberheide and Miller to be vulnerable to detection avoidance by well-crafted malicious apps [1]. Android malware have been found in both official and third party stores. For example, *DroidDream* was distributed through the official Android Market and according to Symantec affected 50,000 to 200,000 users. The third party app stores that have emerged in recent years have also become a very potent source of malicious app distribution as these stores have weak to non-existent measures to prevent malicious apps from being uploaded and distributed to users' devices.

Studies such as [2] have revealed that current families of Android malware are difficult to promptly spot in the wild. This is because of the evasion techniques being used to conceal malicious payload, usually within seemingly innocuous apps that provide functionalities that users want. By employing polymorphic techniques and encrypting malicious payload, signature-based scanning is easily bypassed. With increased code obfuscation, malware analysts take longer to uncover the malicious behavior, classify samples, and generate signatures for detecting the new threats. Moreover, some Android malware families like *AnserverBot* are known to have the capability to fetch and execute malicious payloads at run time thus rendering the zero-day detection of such malware by prior signatures quite ineffective.

These challenges call for new and more effective detection approaches to mitigate the impact of evolving Android malware. Hence, in this paper we propose a method for early detection of Android malware by means of parallel machine learning classifiers that utilize diverse algorithms with inherently different characteristics. A number of static app features are used in the learning phase of the model development. The trained models are combined using various combination schemes to yield a composite model that produces a verdict of 'suspicious' or 'benign' in order to classify a given new application.

Utilizing diverse machine learning classifiers in parallel for Android malware detection, has potential benefits beyond accuracy improvement. It is possible to harness the various strengths of the constituent classifiers in other ways such as complementing white box analysis through close observation of intermediate output from the more interpretable base models.

The main contributions of this paper are the following:

- A new Android malware detection approach is developed using parallel machine learning classifiers. To the best of our knowledge this is the first paper to investigate such an approach for proactive Android malware detection.

- Extensive empirical evaluation of the approach by means of real malware samples and benign applications, demonstrating its real-world applicability and capacity for improved detection accuracy.

The rest of the paper is organized as follows: Section II briefly describes Android application structure and how it provides the app features that underpin our machine learning based malware detection approach. Section III presents the machine learning classification algorithms used to develop the composite classification schemes. Section IV describes the methodology employed to investigate and evaluate the performance of the proposed malware detection approach. Section V presents the evaluation results and discussions. Section VI discusses related works. Finally, section VII draws conclusions and highlights our future work.

## II. ANDROID APPLICATION STRUCTURE AND FEATURE EXTRACTION

### A. Android application basics

An Android application (app) is built from four different types of components: Activities, Services, Broadcast Receivers, and Content Providers. Activities are the components that provide GUI functionality to enable user interactivity, whilst Services and Broadcast Receivers operate in the background when an app is running. Content providers encapsulate data to provide to an app via an interface. Many Android apps consist of at least a number of Activities that are invoked via *intents*, whilst the other three building blocks may optionally be present depending on the app's functionality.

Android apps are written in Java and compiled into a single archive file (Android package or APK), along with data and resource files. Android-powered devices use this APK to install the application. An APK consists of several components including: (1) an XML manifest file containing information such as app description, components declaration (i.e. Activities, Services, Broadcast Receivers etc.), and permissions. (2) A *Classes.dex* file that is a Dalvik executable file that runs in its own instance of a Dalvik Virtual Machine. (3) A */res* directory for indexed resources like icons, images, music etc. (4) A */lib* directory for compiled code. (5) */META-INF* folder holding the app certificate and list of resources, SHA-1 digest etc. (6) *Resources.arsc* which is a compiled resource file.

### B. App feature extraction for machine learning

The malware detection approach investigated in this paper utilizes the static features of an Android app extracted from the APK file in order to determine whether it contains malicious payload or not. It therefore relies on a trained composite classification model (described later) to arrive at the decision. The features employed in training the machine learning model are extracted from a corpus of malware samples and benign apps using a bespoke APK analysis tool written in Java. Details of the APK analysis tool can be found in our previous work [3].

Three categories of features are used for the learning phase: 1) *API related features* 2) *App permissions* 3) *Standard OS and Android framework commands*. The API related features are obtained by mining the *Classes.dex* file using the steps described in [3]. They consists of keywords which enable detection of the use of selected standard Android API calls (through which the app interacts with various device functionalities) as well as selected Java API calls used to enrich apps' functionality.

The permission features are keywords that map onto the standard Android permissions which are declared in the manifest file, therefore enabling the extraction of the permissions being requested by the app for its functionalities. For example SEND_SMS keyword allows for the detection of the permission request for the app to be able to send SMS messages, if declared in the manifest file. The commands related features are keywords that detect the presence of Linux commands such as '*chown*', '*mount*' etc. or certain parameters which might be used with these commands. These commands are usually embedded in hidden files within the APK and invoked by rogue apps to enable unusual activities like privilege escalation, launching hidden scripts or embedded malicious binary files, or concealment of malicious activities. Table I presents an overview of the features under their various categories.

TABLE I. OVERVIEW OF FEATURES EXTRACTED FROM THE APPS AND THEIR BROAD CATEGORIES

| Type | Features (keywords) |
|---|---|
| API calls related | abortBroadcast; getDeviceId; getSubscriberId; getCallState;getSimSerialNumber; getLineNumber; getSimCountryIso; getNetworkOperator; getSimOperator; getPackageManager; Runtime.exec(); android.provider.Contacts; android.provider.ContactsContract; HttpPost_init; HttpGet_init; HttpUriRequest; SMSReceiver; bindService; onActivityResult; SecretKey;KeySpec; FindClass; createSubprocess; Ljavax_crypto_Cipher; Ljavax_crypto_spec_Secret; DexClassLoader; sendMultipartTextMessage; Ljava_net_URLDecoder; native; System.loadLibrary; reflectgetClass; getMethod; registerReceiver; intent.action.BOOT_COMPLETED; intent.action.RUN |
| Command related | mount; remount; chmod; chown; /res; /system/bin; /system/bin/sh; /system/app; .jar; .apk; pmsetInstallLocation; pminstall; GET_META_DATA; GET_RECEIVERS; GET_SERVICES; GET_SIGNATURES; GET_PERMISSIONS |
| Permissions | ACCESS_COARSE_LOCATION; ACCESS_FINE_LOCATION; WRITE_SMS; SEND_SMS; WRITE_CALL_LOG; WRITE_APN_SETTINGS; BROADCAST_SMS; RECEIVE_BOOT_COMPLETED; RECEIVE_MMS; RECEIVE_SMS; RECEIVE_WAP_PUSH; RECORD_AUDIO; CALL_PHONE; WRITE_EXTERNAL_STORAGE; CHANGE_WIFI_STATE; CLEAR_APP_CACHE; INSTALL_PACKAGES; INTERNET; CAMERA;CHANGE_CONFIGURATION; CHANGE_NETWORK_STATE [1] |

[1] A total of 125 permissions are used.

## III. MACHINE LEARNING MODELS FOR THE PARALLEL CLASSIFICATION APPROACH

Machine learning (ML) classifiers have played a part in the development of intelligent systems for several domains over the years. ML approaches are gaining traction in identification and detection of malware on both mobile and PC platforms. Our work is based on *supervised machine learning* whereby the features described in the previous section are acquired from a labelled dataset and used to build and train a model. The ML algorithms considered in our investigation include: Decision Tree (tree-based), Simple Logistic (function-based), Naïve Bayes (probabilistic), PART (rule-based), and RIDOR (rule-based).

Our proposed approach in this paper for a machine learning based zero-day Android malware detection is a composite model of the aforementioned heterogeneous classifiers utilized in various parallel combination schemes. The approach is intended to leverage the strengths of different kinds of supervised learning algorithms to produce a single classification verdict for new applications. Hence, the composite model is built from a function-based, tree-based, probabilistic, and two rule based algorithms. Figure 1 illustrates the building blocks of the detection approach. The rule based classifiers produce the most easily interpretable output whilst the probabilistic classifier is most easily amenable to post-training sensitivity tuning.

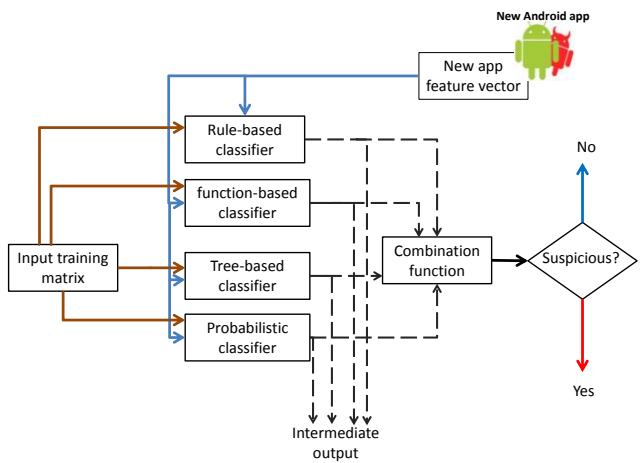

Fig. 1. Android malware detection with the composite parallel classifier approach.

The constituent ML algorithms of the parallel detector include:

*Decision Tree (DT)*: Decision trees are generally known as 'divide and conquer' algorithms. Decision trees are sequential models which logically combine a sequence of simple tests where a numerical attribute (feature) is compared against a threshold value or a nominal attribute (feature) against a set of possible values [4]. It is essentially a flow chart like structure where each internal node denotes a test on an attribute with each branch representing an outcome of the test and each leaf holding a class label.

*Simple Logistic (SL)*: is an ensemble learning algorithm which utilizes additive logistic regression using simple regression functions as base learners [5]. Similar to linear regression, it tries to find a function that will fit the training data well by computing the weights that maximizes the log-likelihood of the logistic regression function. SL classifier takes relatively longer to train but is fast in classification.

*Naïve Bayes (NB)*: The Naïve Bayes classifier operates on the (naïve) assumption of independence of all the features. Despite this simplifying assumption, NB learners and classifiers perform quite well in many real-life applications, most famously in document classification and spam filtering. Compared to more sophisticated methods, NB learning and classification can be extremely fast.

*PART*: is a 'separate-and-conquer' rule learner which produces ordered sets of rules called 'decision lists' [6]. Features from a new app will be compared to each rule in the list in turn, and the app is assigned the category of the first matching rule (a default is applied if no rule successfully matches). PART builds a partial C4.5 decision tree in each iteration and then turns the "best" leaf into a rule.

*RIDOR*: Ridor (Ripple Down Rule learner) [7] is a rule learning algorithm that generates a default rule and then exceptions to the default rule with the least weighted error rate. It generates the best exception for each exception and iterates until pure thus performing a tree-like expansion of exceptions. The exceptions are the rules that predict classes other than the default. Ridor also falls under the general class of 'separate and conquer' ML algorithms.

## IV. INVESTIGATION METHODOLOGY

### A. Input preprocessing

By means of our bespoke APK analysis tool described in [2], the features depicted in Table 1, were extracted and preprocessed into a matrix of input vectors for training the machine learning algorithms. Each column of the matrix represented a single feature, while the rows represented the feature vectors from a single app from the training corpus. The feature vectors were made up of (one of possible values of) 1's and 0's depicting the presence or absence of the corresponding column feature respectively, as detected by the APK analysis tool. A total of 179 training features were extracted, with the breakdown as follows: API calls and commands related: 54 features. App permissions: 125.

### B. Model training

Since supervised learning is the underlying method being used, the training set consisted of app samples labelled in one of two classes; suspicious or benign. A total of 6,863 applications (from McAfee internal repository) were utilized; 2925 malicious apps and 3,938 non-malicious apps. Thus, the input training matrix is of size 6,836 by 180 (179 features and 1 column with the class label).

### C. Model evaluation

In order to evaluate the performance of the classifier model, the 10-fold cross validation technique is applied to the matrix. Thus, the dataset is partitioned into 10 equal parts, k1, k2, k3 to k10 without overlaps. Each step in the evaluation takes one partition as test data and applies a trained model from the other 9 parts. The results are averaged to provide a final performance results for the classifier. k-fold cross validation technique is a very popular ML evaluation method and is appropriate to our goal of determining the relative effectiveness of the ML classifiers in *detecting unknown malicious* apps which is emulated by the non-overlapping testing partitions.

The following performance metrics are used to investigate the parallel classifier approach to Android malware detection.

- *True positive ratio (TPR)*: This is the ratio of correctly classified malicious apps to the total number of malicious apps in the dataset.

- *True negative ratio (TNR)*: The ratio of correctly classified benign apps to the total number of benign apps in the dataset.

- *False positive ratio (FPR)*: The ratio of incorrectly classified benign apps to the total number of benign apps in the dataset.

- *False negative ratio (FNR)*: The ratio of incorrectly classified malicious apps to the total number of malicious apps in the dataset.

- *Accuracy (ACC)*: This is the total accuracy of the classifier given by (TPR + TNR)/ (TPR + TNR + FPR + FNR).

- *Error ratio (ERR)*: This is computed from: 1-ACC.

- *AUC (Area under ROC)*: ROC is the receiver operation characteristics curve. AUC is an estimate of the area under ROC which indicates the predictive power of the classifier. Classifiers with higher AUC have better predictive power and can provide for better sensitivity tuning.

## V. RESULTS AND DISCUSSIONS

### A. Experiment 1: Individual classifiers experiments

In order to obtain baseline results for investigating the parallel classifiers approach to Android malware detection, the first set of experiments were performed with each of the individual candidate classification algorithms for the composite model. These include the Naïve Bayes, PART, RIDOR, Decision Tree and Simple Logistic discussed in section III.

The results from each of these classifiers are summarized together in Table II. The preprocessed input training matrix and the training-testing methodology described in section IV were applied in turn to each classifier to obtain the results presented in Table II.

TABLE II. PERFORMANCE RESULTS FROM THE 5 INDIVIDUAL CLASSIFIERS.

| Algorithm | Performance metrics | | | | | | |
|---|---|---|---|---|---|---|---|
| | TPR | TNR | FPR | FNR | ACC | ERR | AUC |
| NB | 0.821 | 0.913 | 0.087 | 0.179 | 0.867 | 0.133 | 0.915 |
| SL | 0.909 | 0.954 | 0.046 | 0.091 | 0.932 | 0.068 | 0.977 |
| DT | 0.948 | 0.960 | 0.040 | 0.052 | 0.954 | 0.046 | 0.964 |
| RIDOR | 0.957 | 0.942 | 0.058 | 0.043 | 0.950 | 0.050 | 0.949 |
| PART | 0.958 | 0.967 | 0.033 | 0.042 | 0.963 | 0.037 | 0.970 |

The NB classifier can be seen to have the least detection ratio (TPR) of all the five classifiers, and also the least overall accuracy. The SL classifier has the next best malware detection capability with about 91% detection ratio whilst DT, RIDOR and PART showed better detection with 94.8%, 95.7% and 95.8% respectively. In terms of the overall accuracy/error rates, PART proved to be the best performer.

### B. Implications for parallel classification

Recall that PART derives its decision rules from partial decision trees by selecting the 'best' leaf from a branch in each iteration. The number of decision rules derived for the PART model built from the full 6863 by 180 matrix training dataset was *74 rules*. Hence, from our proposed 179 features, a compact rule set is built with PART which is small enough to be used in a parallel classification scenario without incurring excessive classification overhead for new applications. With the RIDOR algorithm, an even smaller rule set of *25 rules* was derived. As for the Decision Tree (J48 algorithm), 143 leaves resulted from model building. This means we have a relatively small decision tree with at most *143 decision branches*. The SL model takes the longest to train since weights have to be determined for each feature in order to fit a logistic regression model to the data. On the other hand, classification of new applications is fast since an already trained model requires only linear additive and multiplicative steps in the decision stage. The NB model is fast to train because it involves calculation of probabilities from the frequencies derived from the input matrix. It is also fast in classification because, like the SL, a decision involves only linear multiplicative and additive computations. Hence, NB and SL are deemed along with the other 3 obtained models, suitable for a combined parallel classification scheme, due to the overall low computational overhead involved in classifying new applications in order to detect the presence of malicious payload.

### C. Experiment 2: parallel classifiers experiments

In the second set of experiments, the combined classification approach which involved a parallel combination of classification decisions obtained from each individual classifier was investigated. Four different combination schemes were considered:

**Average of probabilities**: i.e. an average of the probabilities of each class (suspicious/benign) from the individual classifiers. Thus, a new application is considered suspicious if: Avg. ($P1_{sus} + P2_{sus} + P3_{sus} + P4_{sus} + P5_{sus}$) > Avg. ($P1_{ben} + P2_{ben} + P3_{ben} + P4_{ben} + P5_{ben}$). Otherwise, it is classed as benign.

**Product of probabilities**: i.e. product of the probabilities of each class (suspicious/benign) from the individual classifiers. Thus, a new application is considered suspicious if: ($P1_{sus} \cdot P2_{sus} \cdot P3_{sus} \cdot P4_{sus} \cdot P5_{sus}$) > ($P1_{ben} \cdot P2_{ben} \cdot P3_{ben} \cdot P4_{ben} \cdot P5_{ben}$). Otherwise, it is classed as benign.

**Maximum probability**: i.e. The maximum probabilities for each class (suspicious/benign) of the probabilities output from individual classifier are compared. Thus, a new application is considered suspicious if: Max ($P1_{sus}, P2_{sus}, P3_{sus}, P4_{sus}, P5_{sus}$) > Max ($P1_{ben}, P2_{ben}, P3_{ben}, P4_{ben}, P5_{ben}$). Otherwise, it is classed as benign.

**Majority vote**: For majority vote, individual class decisions are made by each classifier. The majority verdict is taking as the final output decision class.

The results for the various combination schemes used in the parallel classification approach are shown in Table III. The detection rate (TPR) of the parallel classifier approach either equaled or performed better than any of the single classifier baseline TPRs obtained in the first set of experiments. With the *maximum probability* scheme, the best detection rate performance of 97.5% is obtained. This is a detection accuracy improvement over the 95.7% that obtains from the best individual classifier performance in the first set of experiments. The improvement equates to about 50 more malware samples detected in the former scheme than the latter. The detection rate improvement comes at the price of a slight increase in the FPR over that of the best performing individual classifier. However, the *products of probabilities* scheme improves the detection rate to 97.3% without incurring an increase in FPR when compared to the performance of the individual classifiers in Table II.

TABLE III. PERFORMANCE RESULTS FROM FOUR PARALLEL COMBINATION SCHEMES UTILIZING THE 5 CLASSIFIERS

| Combination | Performance metrics | | | | | | |
|---|---|---|---|---|---|---|---|
| | TPR | TNR | FPR | FNR | ACC | ERR | AUC |
| AvgProb | 0.957 | 0.969 | 0.031 | 0.043 | 0.963 | 0.037 | 0.988 |
| ProdProb | **0.973** | **0.970** | **0.030** | **0.027** | **0.972** | **0.028** | **0.953** |
| MaxProb | **0.975** | **0.928** | **0.072** | **0.025** | **0.952** | **0.048** | **0.986** |
| MVote | 0.957 | 0.969 | 0.031 | 0.043 | 0.963 | 0.037 | 0.963 |

From Table III, it can be seen that the best accuracy and TNR results come from the *products of probabilities* combination schemes. All the identification, error/accuracy results from the *products of probabilities* parallel classifiers scheme are better than the baseline performance results from the individual classifiers in the previous set of experiments. The results in Table III demonstrate the efficacy of applying parallel classifiers to detection of Android malware using the features and approach described in this paper. Considering the relatively low classification overhead that the selected diverse base classifiers present, we consider the proposed classification approach a practically viable means of improving Android malware detection to complement existing solutions. Especially, for detecting zero-day Android malware for which no signatures have been derived.

## VI. RELATED WORK

The two main approaches applicable to malware analysis are dynamic analysis and static analysis. Most existing research aimed at non-signature based detection of Android malware generally utilizes either a dynamic or static analysis approach. A few exceptions like the AAS sandbox presented in [8] combine both approaches. Android malware detection approaches based on dynamic analysis can be found in Crowdroid proposed by Burguera et al. in [9]. Crowdroid is a behavior based malware detection system for Android that uses run-time system call features and clustering algorithms to detect Android malware. MADAM [10] is also a dynamic analysis based anomaly detector for Android malware. MADAM monitors Android at the kernel-level and user-level and applies machine learning classifiers. MADAM was tested on 10 monitored real malware and according to the authors, showed a negligible impact on the user experience.

In [11], Shabtai et al. proposed Andromaly, a host-based Android malware detection solution that employs dynamic analysis. Andromaly continuously monitors various features and events like CPU consumption, number of packets sent, number of running processes, keyboard/touch-screen pressing etc. Machine learning anomaly detectors are then applied to classify the collected data into normal or abnormal. In [12], AntiMalDroid was proposed by Zhao et al. AntiMalDroid is a dynamic analysis behavior based malware detection framework that uses logged behavior sequence as features for SVM model training and detection.

Different from these previous works that are based on dynamic analysis, the approach in this paper employs a static analysis based approach for malware detection. In this case, static code properties are used to proactively identify malware before it is installed and run on a device. Hence our method can be applied to screening a large number of apps on an app market in a relatively short time period. Furthermore, the resource constraints imposed by the handheld devices are avoided by our approach. Another important advantage of a static based approach over a dynamic one is that it is undetectable by the malware itself i.e. malware cannot modify its behavior during analysis [2]. Static analysis offers faster, less resource intensive and more code coverage in less time than dynamic runtime analysis.

Previous Android malware analysis works which employ static analysis include DroidMat [13]. DroidMat uses k-means clustering to detect malware based on static functional behaviors derived from API and permissions detected in the application. In [14], a static analysis based Bayesian classification method was developed to categorize apps into 'benign' or 'suspicious' using 58 static code-based features. The training and classification employed 1000 Android malware samples from 49 families and 1000 benign applications. The approach in [15] utilized permissions and call flow graphs for training SVM models to distinguish between benign and malicious Android apps. The authors derived one-class SVM models based on the benign samples alone and use these for identification of both benign and malicious apps. In [16], the authors also apply machine learning with static analysis, but utilize Linux malware rather than Android malware samples. Their approach extracts Linux system commands within Android and use the readelf command to output a list of referenced function calls for each system command. The same method is used to extract a static list of function calls with 240 Linux virus, worms, and Trojans. Both sets are applied to train PART, Prism and Nearest Neighbor algorithms for classification.

In [17], Sanz et al compared various machine learning schemes trained with permission features on their malware detection accuracy. Their analysis was based on 249 malware samples and 347 benign apps. Sarma et al. [18] and Peng et al. [19] also apply permissions to train SVM based and Bayesian based models respectively for risk ranking of Android apps. The study in this paper also utilizes permissions as features but differs from the previous by including a more extensive feature set not used in the previous works. For example, command related features are not used in previous works (except in [14]). From our experience, these features have been found to be quite effective in enhancing the classification accuracy of trained machine learning models that are based on static analysis. Moreover, this paper proposes and evaluates a more effective way of leveraging static code features for Android malware through parallel machine learning classification schemes. The research in [2] and [20] are some of the previous work that are not based on machine learning approaches but do apply static analysis for Android malware detection.

## VII. CONCLUSION

In this paper a parallel classification approach to Android malware detection using inherently diverse machine learning algorithms was investigated. The proposed approach utilized a wide range of features which included API calls related, commands related and permission features. The recent increase in Android malware and their growing ability for adept detection avoidance of existing signature-based approaches definitely calls for novel alternatives. The parallel classification approach proposed in this paper is a viable scheme that provides a complementary tool that not only potentially improves Android malware detection but also allows the strengths of diverse classifiers to be leveraged. For example, the rule based classifiers can provide human-interpretable intermediate output that can be useful for driving further analysis stages. Furthermore, the proposed approach is ideal from performance point of view since it is cost effective in classifying a new application because: 1) static app features are employed and 2) the selected constituent classification models have low computational requirements during classification decision.

As future work, we aim to develop and evaluate an Android malware detection engine using the investigated approach. Further study involving the performance tuning of the detection engine when applied to new datasets from emerging Android app markets will also be considered.

ACKNOWLEDGMENT

We gratefully acknowledge McAfee's support in providing the repository of malware and benign apps used for this research work.